\documentclass[a4paper,11pt]{article}
\usepackage{stmaryrd}
\usepackage{amsfonts}
\usepackage{bbm}
\usepackage{float}
\usepackage{amscd}
\usepackage{mathrsfs}
\usepackage{latexsym,amssymb,amsmath,amscd,amscd,amsthm,amsxtra,xypic}
\usepackage[dvips]{graphicx}
\usepackage[utf8]{inputenc}
\usepackage[T1]{fontenc}
\usepackage{lmodern}
\usepackage{amssymb}
\usepackage[all]{xy}
\usepackage{nicefrac,mathtools,enumitem}
\usepackage{microtype}
\usepackage{lipsum}
\usepackage{mathtools}
\usepackage{xfrac}

\usepackage{geometry}
\usepackage{hyperref}

\usepackage{times}
\usepackage{amsthm}
\usepackage{amsmath}
\usepackage{amsfonts}
\usepackage{amssymb}
\usepackage{mathdots}
\usepackage{color}
\usepackage{mathrsfs}
\usepackage{stmaryrd}
\SetSymbolFont{stmry}{bold}{U}{stmry}{m}{n}

\usepackage{bm}
\usepackage{tikz-cd}
\usepackage{tikz}
\usetikzlibrary{matrix,arrows,decorations.pathmorphing}
\usepackage[all]{xy}
\usepackage{graphicx}
\usepackage{subfigure}
\usepackage[toc,page]{appendix}
\usepackage[usestackEOL]{stackengine}

\usepackage{theoremref}
\newtheorem{thm}{Theorem}[section]
\newtheorem{lem}[thm]{Lemma}

\newtheorem{pro}[thm]{Proposition}

\theoremstyle{definition}

\newtheorem{rmk}[thm]{Remark}
\newtheorem{defi}[thm]{Definition}

\newcommand{\be }{\begin{equation}}
\newcommand{\ee }{\end{equation}}

\newcommand{\pf}{\noindent{\bf Proof.}\ }

\newcommand{\h}{\mathfrak h}

\def\qed{\hfill ~\vrule height6pt width6pt depth0pt}

\newcommand{\br}[1]{   [ \cdot,    \cdot  ]   }

\newcommand{\gl}{\mathfrak {gl}}

\newcommand{\Herm}{{\rm Herm}}

\title{On the connection formula of a higher rank analog of Painlev\'{e} VI}

\author{Xiaomeng Xu}
\date{}

\newcommand{\Addresses}{{
  \bigskip
  \footnotesize
\noindent \textsc{
School of Mathematical Sciences \& Beijing International Center
for Mathematical Research, Peking University, Beijing 100871, China}\par\nopagebreak
  \textit{E-mail address}: \texttt{xxu@bicmr.pku.edu.cn}
}}
\begin{document}

\maketitle
    
\begin{abstract}
In this paper, we derive the connection formula for the higher rank analog of Painlev\'{e} VI arsing from the isomonodromic deformation theory and the study of Frobenius manifolds.
\end{abstract}
 
\section{Introduction}
The six classical Painlev\'{e} equations were introduced at the turn of the twentieth century by Painlev\'{e} \cite{Pa} and Gambier \cite{Ga}, in a specific
classification problem for second order ODEs. Since then, they have appeared in the integrable nonlinear PDEs, 2D Ising models, random matrices, topological field theory and so on. 
We refer the reader to the book of Fokas, Its, Kapaev and Novokshenov \cite{FIKN} for a thorough introduction to the history and developments of the study of Painlev\'{e} equations. As stressed in \cite{FIKN,IN}, the solutions of Painlev\'{e} equations (called Painlev\'{e} transcendents) are seen as nonlinear special functions, because they play the same role in nonlinear mathematical physics as that of classical special functions, like Airy functions, Bessel functions, etc., in linear physics. And it is the answers of the following questions $(a)$ and $(b)$ make Painlev\'{e} transcendents as efficient in applications as linear special functions (see \cite{FIKN, IN} and the references therein for more details):
\begin{itemize}
    \item[(a).] The parametrization of Painlev\'{e} transcendents by their asymptotic behaviour at critical points;
     
\item[(b).] The construction of the connection 
formula from one critical point to another.

\end{itemize}
Many global asymptotic results, including the connection formula, of Painlev\'{e} transcendents are based on the method of isomonodromic deformation, introduced by Flaschka-Newell \cite{FN}, and by Jimbo-Miwa-Ueno \cite{JMU}.

The theory of isomonodromic deformation, as well as the theory of Frobenius manifolds (see the book \cite{Dubrovin} of Dubrovin), leads to a natural generalization of a higher rank analog of the sixth Painlev\'{e} equation. Let $\h_{\rm reg}$ denote the set of $n\times n$ diagonal matrices with distinct eigenvalues. Then, the higher rank analog of the sixth Painlev\'{e} equation is the nonlinear differential equation for a $n\times n$ matrix valued function $\Phi(u): \h_{\rm reg}\rightarrow \frak{gl}(n)$
\begin{eqnarray}\label{isoeq}
\frac{\partial \Phi}{\partial u_k} =\frac{1}{2\pi\iota}[\Phi,{\rm ad}^{-1}_u{\rm ad}_{E_{kk}}\Phi], \ \text{for all} \ k=1,...,n.\end{eqnarray} 
Here the square brackets take the matrix commutator, $u={\rm diag}(u_1,...,u_n)\in\h_{\rm reg}$, and $\iota:=\sqrt{-1}$, $E_{kk}$ is the $n\times n$ diagonal matrix whose $(k,k)$-entry is $1$ and other entries are $0$.
Note that the adjoint ${\rm ad}_{E_{kk}}\Phi=[E_{kk},\Phi]$ takes values in the subspace of off-diagonal matrices, and that the adjoint operator ${\rm ad}_u$
is invertible when restricted to the subspace. In this sense ${\rm ad}^{-1}_u{\rm ad}_{E_{kk}}\Phi$ denotes the unique off-diagonal matrix such that $[u,{\rm ad}^{-1}_u{\rm ad}_{E_{kk}}\Phi]={\rm ad}_{E_{kk}}\Phi$.

The equation \eqref{isoeq} is the isomonodromic deformation equation of meromorphic linear systems of ordinary differential equations with Poncar\'{e} rank $1$. See Section \ref{sec:iso}. Following Miwa \cite{Miwa}, the solutions of the equation \eqref{isoeq} have the Painlev\'{e} property. Furthermore, they are multi-valued meromorphic functions of $u_1,...,u_n$ and the branching occurs when $u$ moves along a loop around the fat diagonal
\[\Delta=\{(u_1,...,u_n)\in \mathbb{C}^n~|~u_i = u_j, \text{for some } i\ne j \}.\]
Thus, according to the original idea of Painlev\'{e}, they can be a new class of special functions. In the meanwhile, these functions have found applications in the theory of Frobenius manifolds \cite{Dubrovin}, solving the Witten-Dijkgraaf-Verlinde-Verlinde (WDVV) equations \cite{Gu0}, stability conditions \cite{BTL}, crystal bases in representation theory \cite{Xu}, the WKB approximation of linear systems of differential equations \cite{AXZ} and so on. However, in the literature there is no
general results of the similar questions $(a)$ and $(b)$ for \eqref{isoeq}.

To fill the gap, in the previous paper \cite{Xu}, we studied the problem $(a)$, i.e., characterized the asymptotics of solutions of the equation \eqref{isoeq} in all asymptotic zones of $u$, using the idea of the Gelfand-Tsetlin theory in representation theory. Furthermore, we obtained the expression of the monodromy of the associated linear problem. We remark that the explicit expression of monodromy data would give some constrains on the distributions of poles of solutions of \eqref{isoeq}. 

In this paper, we use the results in \cite{Xu} to derive an answer to the problem $(b)$, i.e., obtaining the explicit connection formula between two asymptotic zones, via the method of isomonodromic deformation. As will become clear below, we thus obtain the global information about the equation \eqref{isoeq} using only a local analysis.

\subsection*{The case $n=3$: Painlev\'{e} VI}
Before stating our main result, let us give a brief recall of the sixth Painlev\'{e} equation (simply denoted by PVI) as an illustrative example. That is the nonlinear differential equation
\begin{eqnarray*}
\frac{d^2y}{dx^2}&=&\frac{1}{2}\Big[\frac{1}{y}+\frac{1}{y-1}+\frac{1}{y-x}\Big](\frac{dy}{dx})^2-\Big[\frac{1}{x}+\frac{1}{x-1}+\frac{1}{y-x}\Big]\frac{dy}{dx}\\
&+&\frac{y(y-1)(y-x)}{x^2(x-1)^2}\Big[\alpha+\beta\frac{x}{y^2}+\gamma\frac{x-1}{(y-1)^2}+\delta\frac{x(x-1)}{(y-x)^2}\Big], \ \alpha,\beta,\gamma,\delta\in\mathbb{C}.
\end{eqnarray*}
A solution $y(x)$ of PVI has $0,1,\infty$ as critical points, and can be analytically
continued to a meromorphic function on the universal covering of $\mathbb{P}^1\setminus\{0, 1,\infty\}$. The function $y(x)$ is in general not given in terms of classical functions. Then, following \cite{FIKN}, to "solve" the Painlev\'{e} equation, it means that

\begin{itemize}
    \item to find the explicit asymptotics of $y(x)$ at the critical points $x=0,1,\infty$:
    \begin{eqnarray}\label{3asymptotics}
    y(x)\sim y_p(x;a_p,\sigma_p), \ \text{as} \ x\rightarrow p, \ p\in\{0,1,\infty\},
    \end{eqnarray}
where $a_p,\sigma_p$ are the asymptotic parameters;

\item to find the explicit connection formula of $y(x)$ between two different critical points $p\ne q\in \{0,1,\infty\}$, i.e., the explicit formula
\begin{eqnarray}\label{3connection}
a_p=a_p(a_q,\sigma_q), \hspace{5mm} \sigma_p=\sigma_p(a_q,\sigma_q).
\end{eqnarray}
\end{itemize} 

The asymptotics \eqref{3asymptotics} and connection formula \eqref{3connection} for $y(x)$ were evaluated via the isomonodromy approach for the generic case by Jimbo \cite{Jimbo}, and for important special cases, not covered by Jimbo's
results, by Dubrovin-Mazzocco \cite{DM} and Guzzetti \cite{Gu}. The key feature is that the parameters $(a_p, \sigma_p)$ for $p=0,1,\infty$ can be expressed as functions
of the monodromy data of the associated linear system of differential equation, then the isomonodromy method enables one to solve the
connection problem.

It was shown in \cite{DG} that Painlev\'{e} VI is equivalent to the equation \eqref{isoeq} for $n=3$ with suitable matrices $\Phi(u)$. In particular, let us assume that there exists parameters $\theta_1,\theta_2,\theta_3, \theta_\infty$
\begin{eqnarray}\label{theta}
{\rm diag}(\Phi(u))&=&2\pi\iota \cdot {\rm diag}(\theta_1,\theta_2,\theta_3), \\
\text{eigenvalues of } \Phi(u) &=&0, \ \pi\iota\cdot (\theta_1+\theta_2+\theta_3-\theta_\infty), \ \pi\iota\cdot(\theta_1+\theta_2+\theta_3+\theta_\infty).
\end{eqnarray}
Then the equation \eqref{isoeq} is equivalent to the Painlev\'{e} VI of $y(x)$ with \[x=\frac{u_2-u_1}{u_3-u_1}\] and the parameters \[2\alpha=(\theta_\infty-1)^2, \ 2\beta=-\theta_1^2, \ 2\gamma=-\theta_3^2, \ 2\delta=-\theta_2^2.\] 
Therefore, the asymptotics \eqref{3asymptotics} and connection problem \eqref{3connection} of Painlev\'{e} VI amounts to the study of the asymptotics of solutions of the equation \ref{isoeq} as 
\begin{eqnarray}\label{threepts}
\frac{u_2-u_1}{u_3-u_1}\rightarrow \infty, \hspace{2mm} \frac{u_3-u_1}{u_2-u_1}\rightarrow \infty, \hspace{2mm} \frac{u_2-u_1}{u_2-u_3}\rightarrow \infty,
\end{eqnarray}
respectively, and the connection problem between these asymptotics. This perspective has been taken in \cite{Gu0} with applications in solving the WDVV equations via the asymptotic properties of the equation \eqref{isoeq} for $n=3$.

\section{Asymptotics of solutions of the equation \eqref{isoeq}}\label{firstsec}
One important observation is that the prescription of the limits in \eqref{threepts} is controlled by the De Concini-Procesi space \cite{dCP} of $\h_{\rm reg}$ with $n=3$. In \cite{Xu} we restrict to the "real slice" and describe the asymptotics of solutions of \eqref{isoeq} for general $n$ via the geometry of the corresponding De Concini-Procesi space, as a generalization of the formula in \eqref{3asymptotics}. In this section, we mainly recall the asymptotics in two special asymptotic zones. 

\subsection{The asymptotic behaviours in two special zones}

Let $\h_{\rm reg}(\mathbb{R})$ denote the set of diagonal matrices with distinct real eigenvalues. Let $\Herm(n)$ denote the space of $n\times n$ Hermitian matrices. In this paper, let us only consider the equation \eqref{isoeq} for a function $\Phi(u):\h_{\rm reg}(\mathbb{R})\rightarrow \Herm(n)$. This reduction is equivalent to a complex conjugation symmetry on the system, which in turn guarantees that $\Phi(u)$ is real analytic on the real part $\h_{\rm reg}(\mathbb{R})$ (see \cite{Xu} for an argument based on the work \cite[Theorem 2]{Boalch} of Boalch), and simplifies the asymptotic analysis. 

\begin{thm}\cite{Xu}\label{isomonopro}
For any solution $\Phi(u)\in \Herm(n)$ of the equation \eqref{isoeq} on the connected component $U_{\rm id}:=\{u\in \h_{\rm reg}(\mathbb{R})~|~u_1<\cdots <u_n\}$, there exists a constant $A_{\infty}\in\Herm(n)$ such that 
\begin{eqnarray}\label{firstasy}
\Phi(u)=\Big(\overrightarrow{\prod_{k=0,...,n-1} }(\frac{u_{k}}{u_{k+1}})^{\frac{\delta_k(A_{\infty})}{2\pi\iota}}\Big)^{-1}\cdot A_{\infty}\cdot \Big(\overrightarrow{\prod_{k=0,...,n-1} }(\frac{u_{k}}{u_{k+1}})^{\frac{\delta_k(A_{\infty})}{2\pi\iota}}\Big)+O\Big(\frac{{\rm log}(u_2-u_1)}{u_2-u_1}\Big), 
\end{eqnarray}
as $\frac{u_k-u_{k-1}}{u_{k+1}-u_k}\rightarrow 0$ for $k=1,...,n-1$, where the product $\overrightarrow{\prod}$ is taken with the index $i$ to the right of $j$ if $i>j$ and $u_{0}:=1$, and $\delta_k(\Phi)$ is the matrix with nonzero entries $\delta_k(\Phi)_{ij}=\Phi_{ij}$, if $1\le i, j\le k$, or $i=j.$ Conversely, given any $A_\infty\in\Herm(n)$, there exists a solution $\Phi(u):U_{\rm id}\rightarrow \Herm(n)$ such that \eqref{firstasy} holds.
\end{thm}
Similarly, we get
\begin{thm}\cite{Xu}
For any solution $\Phi(u)\in\Herm(n)$ of the equation \eqref{isoeq} on the connected component $U_{\rm id}$, there exists a constant $A_{-\infty}\in\Herm(n)$ such that 
\begin{eqnarray}\label{secondasy}
\Phi(u)=\Big(\overrightarrow{\prod_{k=1,...,n} }(\frac{u_{k}}{u_{k+1}})^{\frac{\eta_k(A_{-\infty})}{2\pi\iota}}\Big)^{-1}\cdot A_{-\infty}\cdot \Big(\overrightarrow{\prod_{k=1,...,n} }(\frac{u_{k}}{u_{k+1}})^{\frac{\eta_k(A_{-\infty})}{2\pi\iota}}\Big)+O\Big(\frac{{\rm log}(u_{n}-u_{n-1})}{u_n-u_{n-1}}\Big)
\end{eqnarray}
as $\frac{u_{k+1}-u_{k}}{u_{k}-u_{k-1}}\rightarrow 0$ for $k=2,...,n$, where the product $\overrightarrow{\prod}$ is taken with the index $i$ to the right of $j$ if $i<j$ and $u_{n+1}:=1$, and $\eta_k(\Phi)$ is the matrix with nonzero entries $\eta_k(\Phi)_{ij}=\Phi_{ij}$, if $n-k+1\le i, j\le n$, or $i=j.$  Conversely, given any $A_{-\infty}\in\Herm(n)$, there exists a solution $\Phi(u):U_{\rm id}\rightarrow \Herm(n)$ such that \eqref{secondasy} holds.
\end{thm}
Let us denote by $u_{\rm cat}^{(+)}$ the asymptotic zone in $U_{\rm id}$ 
\begin{eqnarray}\label{zone+}
\frac{u_k-u_{k-1}}{u_{k+1}-u_k}\rightarrow 0,\hspace{3mm} for \ k=1,...,n-1,
\end{eqnarray}
and by $u_{\rm cat}^{(-)}$ the zone  \begin{eqnarray}\label{zone-}
\frac{u_{k+1}-u_k}{u_k-u_{k-1}}\rightarrow 0,\hspace{3mm} for \ k=2,...,n.
\end{eqnarray}
Then the identities \eqref{firstasy} and \eqref{secondasy} take the form of the expression \eqref{3asymptotics} at $u_{\rm cat}^{(+)}$ and $u_{\rm cat}^{(-)}$, with the matrices $A_\infty$ and $A_{-\infty}$ as the asymptotics parameters ("initial conditions") respectively. We stress that each of the parameter matrices completely characterizes the solution $\Phi(u)$. 

Now the connection problem, between $u_{\rm cat}^{(+)}$ and $u_{\rm cat}^{(-)}$, is to express the asymptotics $A_\infty$ of any solution $\Phi(u)$ as a function of the asymptotics $A_{-\infty}$ of the same solution.

\subsection{The asymptotic behaviours in generic asymptotic zones}
We have seen an answer to the questioin $(a)$ for the equation \eqref{isoeq} in the asymptotic zones $u_{\rm cat}^{(+)}$ and $u_{\rm cat}^{(-)}$. There are many other asymptotic zones. For example, one typical asymptotic zone for $n=6$ can be described by the rooted planar binary tree

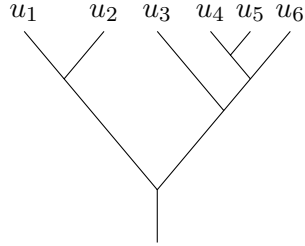
\begin{figure}[H]
\begin{center}
  \begin{tikzpicture}[scale=0.7]
  \draw
  (0,0)--(0,1)--(2.5,4) node[above]{$u_6$}
  (0,1)--(-2.5,4) node[above]{$u_1$}
  (1.25,2.5)--(0,4) node[above]{$u_3$}
  (1.75,3.1)--(1,4) node[above]{$u_4$}
  (1.375,3.55)--(1.75,4) node[above]{$u_5$}
  %(-2.125,3.55)--(-1.75,4) node[above]{1}
  (-1.75,3.1)--(-1,4) node[above]{$u_2$};
  \end{tikzpicture}
  \caption{A rooted planar binary tree with $6$ leaves labelled by $u_1,...,u_6$.}
 \end{center}
 \end{figure}
The tree represents the asymptotic zone where $\frac{u_i-u_j}{u_k-u_l}\rightarrow 0$, if $u_i$ and $u_j$ are the labelling of two leaves from the two branches of an inner vertex $I$, while $u_k$ and $u_l$ are the labelling of two leaves from the two branches of a preceding vertex of $I$. For example, we have $\frac{u_4-u_5}{u_5-u_6}\rightarrow 0$, $\frac{u_4-u_6}{u_3-u_4}\rightarrow 0$, $\frac{u_1-u_2}{u_2-u_3}\rightarrow 0$ and so on. In general (for any dimension $n$), all such asymptotic zones are characterized by planar binary trees (or equivalently correspond to the $0$-dimensional stratum of the the De Concini-Procesi space of $\h_{\rm reg}(\mathbb{R})$). 

The asymptotic behaviour of the solutions of the isomonodormy equation \eqref{isoeq} as $u$ in any asymptotic zone $u_{\rm asy}$ has been studied in \cite{Xu}. Since it takes a rather similar form to the asymptotics at $u_{\rm cat}^{(+)}$ given in Theorem \ref{isomonopro}, we ignore its concrete expression here and refer the reader to \cite{Xu} for details.

\section{Connection formula via the method of isomonodromic deformation}
The nonlinear differential equation \eqref{isoeq} governs the isomonodromic deformation of a meromorphic linear system of ordinary differential equations. Furthermore, the monodromy data of the linear system is a complete set of independent first integrals of \eqref{isoeq}. It thus transfers global analytic problems of equation \eqref{isoeq} to the study of the monodromy of the linear system (we refer the reader to the book \cite{IN} for an introduction to the isomonodromic deformation method in the study of Painlev\'{e} equations). The key feature of the associated linear system is the appearance of irregular singularities, which involves the study of Stokes phenomenon. Thus, an analytic basis 
for the achievement of the connection formula consists of solving explicitly the Riemann-Hilbert-Birkhoff problem of the linear system in asymptotic zones (see Theorem \ref{mainthm} for one example, and see \cite{Xu} for more details).

This section applies the idea of isomonodromic deformation to solve the connection problem for \eqref{isoeq}. 
Section \ref{sec:Stokes} introduces the involved meromorphic linear system of ordinary differential equations and recalls the notion of its Stokes matrices. Section \ref{sec:iso} explains that in which sense the equation \eqref{isoeq} is the isomonodromic deformation equation of the linear system. Section \ref{expStokes} explicitly expresses the Stokes matrices of the linear system via the asymptotics of solutions of \eqref{isoeq}. In the end, Section \ref{conntwozones} gives the explicit connection formula between the two asymptotic zones $u_{\rm cat}^{(+)}$ to $u_{\rm cat}^{(-)}$, and the last subsection briefly discusses the connection formula between all asymptotic zones.

\subsection{Meromorphic linear system of Poncar\'e rank 1 and Stokes matrices}\label{sec:Stokes}
Let us consider the meromorphic linear system
\begin{eqnarray}\label{Stokeseq}
\frac{dF}{dz}=\Big(\iota u-\frac{1}{2\pi\iota}\frac{A}{z}\Big)\cdot F,
\end{eqnarray}
where $F(z)$ is valued in ${\rm GL}_n(\mathbb{C})$, $u\in\h_{\rm reg}(\mathbb{R})$ and $A\in{\Herm}(n)$. The system has an order two pole at $\infty$ and (if $A\neq 0$) an order one pole at $0$. 

\begin{defi}\label{Stokesrays}
The {\it Stokes sectors} of the system
are the right/left half planes ${\rm Sect}_\pm=\{z\in\mathbb{C}~|~ \pm{\rm Re}(z)>0\}$.
\end{defi}
For any two real numbers $a, b$, an open sector and a closed sector with opening
angle $b-a>0$ are respectively denoted by
\[S(a,b):=\{z\in\mathbb{C}~|~a<{\rm arg}(z)<b\}, \hspace{5mm} \bar{S}(a,b):=\{z\in\mathbb{C}~|~a\le {\rm arg}(z)\le b\}. \]

Let us choose the branch of ${\rm log}(z)$, which is real on the positive real axis, with a cut along the nonnegative imaginary axis $\iota \mathbb{R}_{\ge 0}$. Then by convention, ${\rm log}(z)$ has imaginary part $-\pi$ on the negative real axis in ${\rm Sect}_-$. The following construction of solutions via the Laplace-Borel transform is standard, see e.g., \cite{Balser, LR, Xu}. 
\begin{thm}\label{uniformresum}
For any $u\in\h_{\rm reg}(\mathbb{R})$, on ${\rm Sect}_\pm$ there
is a unique (therefore canonical) fundamental solution $F_\pm:{\rm Sect}_\pm\to {\rm GL}_n(\mathbb{C})$ of equation \eqref{Stokeseq} such that $F_+\cdot e^{-\iota uz}\cdot z^{\frac{[A]}{2\pi\iota}}$ and $F_-\cdot e^{-\iota uz}\cdot z^{\frac{[A]}{2\pi\iota}}$ can be analytically continued to $S(-\pi,\pi)$ and $S(-2\pi,0)$ respectively, and for every small $\varepsilon>0$,
\begin{eqnarray*}
\lim_{z\rightarrow\infty}F_+(z;u)\cdot e^{-\iota uz}\cdot z^{\frac{[A]}{2\pi\iota}}&=&{\rm Id}_n, \ \ \ as \ \ \ z\in \bar{S}(-\pi+\varepsilon,\pi-\varepsilon),
\\
\lim_{z\rightarrow\infty}F_-(z;u)\cdot e^{-\iota uz}\cdot z^{\frac{[A]}{2\pi\iota}}&=&{\rm Id}_n, \ \ \ as \ \ \ z\in \bar{S}(-2\pi+\varepsilon,-\varepsilon),
\end{eqnarray*}
Here ${\rm Id}_n$ is the rank n identity matrix, and $[A]$ is the diagonal part of $A$. The solutions $F_\pm$ are called the canonical solutions in ${\rm Sect}_\pm$.
\end{thm}

For any $\sigma\in S_n$, let us denote by $U_{\sigma}$ the connected component \[U_{\sigma}:=\{u={\rm diag}(u_1, . . . , u_n)\in \h_{\rm reg}(\mathbb{R})~|~ u_{\sigma(1)}<\cdot\cdot\cdot < u_{\sigma(n)}\}\] of $\h_{\rm reg}(\mathbb{R})$, and denote by $P_\sigma\in{\rm GL}_n(\mathbb{C})$ the corresponding permutation matrix.

\begin{defi}\label{defiStokes}
For any $u\in U_\sigma$, the {\it Stokes matrices} of the system \eqref{Stokeseq} (with respect
to ${\rm Sect}_+$ and the chosen branch of ${\rm log}(z)$) are the elements $S_\pm(u,A)\in {\rm GL}_n(\mathbb{C})$ determined by
\[F_+(z)=F_-(z)\cdot e^{-\frac{[A]}{2}}P_\sigma S_+(u,A)P_\sigma^{-1}, \  \ \ \ \ 
F_-(ze^{-2\pi \iota})=F_+(z)\cdot P_\sigma S_-(u,A)P_\sigma^{-1}e^{\frac{[A]}{2}},
\]
where the first (resp. second) identity is understood to hold in ${\rm Sect}_-$
(resp. ${\rm Sect}_+$) after $ F_+$ (resp. $F_-$)
has been analytically continued anticlockwise around $z=\infty$. 
\end{defi} 
According to our definition, the (diagonal part of) Stokes matrices include the information of the formal monodromy data of \eqref{Stokeseq}. It is slightly different from the usual convention in the literature (see e.g., \cite{Boalch} where Stokes matrices are defined such that all diagonal entries are $1$, and the formal monodromy is seen an extra data). 

The prescribed asymptotics of $F_\pm(z)$ at $z=\infty$, as well as the identities in Definition \ref{defiStokes}, ensures that the Stokes matrices $S_+(u,A)$ and $S_-(u,A)$ are upper and lower triangular matrices respectively. see e.g., \cite[Lemma 17]{Boalch} or \cite[Chapter 9.1]{Balser}. Furthermore, the following lemma follows from the fact that if $F(z)$ is a solution, so is the complex conjugation $F(\bar{z})^\dagger$ of the matrix function $F(\bar{z})$, see \cite[Lemma 29]{Boalch}. It reflects the "real condition" imposed on the coefficient of \eqref{Stokeseq}.
\begin{lem}
Let $S_+(u,A)^\dagger$ denote the conjugation transpose of $S_+(u,A)$, then $S_-(u,A)=S_+(u,A)^\dagger$.
\end{lem}

\subsection{Isomonodromic deformation}\label{sec:iso}
In general, the Stokes matrices $S_\pm(u,A)$ of the system \eqref{Stokeseq} will depend on the irregular term $u$. The isomonodromic deformation (also known as monodromy preserving) problem is to find the matrix valued function $A(u)$ such that the Stokes matrices $S_\pm(u,A(u))$ are (locally) constant. In particular, the following proposition is well known. See more detailed discussions in e.g., \cite{BoalchG,CDG, Dubrovin,JMU}.
\begin{pro}\label{isomonodef}
For any solution $\Phi(u)$ of the equation \eqref{isoeq}, the Stokes matrices $S_\pm(u,\Phi(u))$
are locally constant (independent of $u$).
\end{pro}
That is one can associate with initial non-linear differential equation \eqref{isoeq} a linear system 
\begin{eqnarray}\label{isoStokeseq}
\frac{dF}{dz}=\Big(\iota u-\frac{1}{2\pi\iota}\frac{\Phi(u)}{z}\Big)\cdot F,
\end{eqnarray}
where the variation of $\Phi(u)$ in the
coefficient is described by equation \eqref{isoeq}. Then the monodromy data (Stokes matrices) of system \eqref{isoStokeseq} is conserved. In this way, \eqref{isoeq} is explained as the isomonodromic deformation equation of the linear system \eqref{isoStokeseq}. In the following, the equation \eqref{isoeq} is also called the isomonodromy equation.

\subsection{Expression of the Stokes matrices $S_\pm(u,\Phi(u))$ via the asymptotics of $\Phi(u)$}\label{expStokes}
In \cite{Xu}, we develop the analytic branching rule of the system \eqref{Stokeseq}. Roughly speaking, it states that if components $u_i$ of $u$ collapse according to the branching of a planar tree, then the Stokes matrices of the system \eqref{Stokeseq} can be explicitly recovered by the Stokes matrices of the lower rank systems determined by the branching of the tree. As an application, it gives an expression of the Stokes matrices of the system \eqref{isoStokeseq} via the asymptotics of solutions of the isomonodromy equation as follows.

\begin{defi}\label{solcat+}
We denote by $\Phi(u;A_{\infty})$ the solution of the isomonodromy equation \eqref{isoeq} with the asymptotics $A_{\infty}$ at $u_{\rm cat}^{(+)}$ in the sense of \eqref{firstasy}.
\end{defi}
We denote by $\lambda^{(k)}_1\le \lambda^{(k)}_2\le\cdots\le \lambda^{(k)}_k$ the ordered eigenvalues of the left-top $k\times k$ submatrix of the Hermitian matrix $A_{\infty}$, and $\lambda^{(k)}_{k+1}=\sum_{i=1}^{k+1}\lambda^{(k+1)}_i-\sum_{i=1}^{k}\lambda^{(k)}_i$. Then
\begin{thm}\cite{Xu}\label{mainthm}
The sub-diagonals of the Stokes matrices $S_\pm(u,\Phi(u;A_\infty))$ are given by
\begin{eqnarray*}
(S_+)_{k,k+1}&=&e^{\frac{\small{\lambda^{(k-1)}_{k}-\lambda^{(k)}_{k+1}}}{4}} \sum_{i=1}^k\frac{\prod_{l=1,l\ne i}^{k}\Gamma(1+\frac{\lambda^{(k)}_l-\lambda^{(k)}_i}{2\pi \iota})}{\prod_{l=1}^{k+1}\Gamma(1+\frac{\lambda^{(k+1)}_l-\lambda^{(k)}_i}{2\pi \iota})}\frac{\prod_{l=1,l\ne i}^{k}\Gamma(1+\frac{\lambda^{(k)}_l-\lambda^{(k)}_i}{2\pi \iota})}{\prod_{l=1}^{k-1}\Gamma(1+\frac{\lambda^{(k-1)}_l-\lambda^{(k)}_i}{2\pi \iota})}\cdot m^{(k)}_i,\\
(S_-)_{k+1,k}&=&e^{\frac{\small{\lambda^{(k-1)}_{k}-\lambda^{(k)}_{k+1}}}{4}}
\sum_{i=1}^k \frac{\prod_{l=1,l\ne i}^{k}\Gamma(1-\frac{\lambda^{(k)}_l-\lambda^{(k)}_i}{2\pi \iota})}{\prod_{l=1}^{k+1}\Gamma(1-\frac{\lambda^{(k+1)}_l-\lambda^{(k)}_i}{2\pi \iota})}\frac{\prod_{l=1,l\ne i}^{k}\Gamma(1-\frac{\lambda^{(k)}_l-\lambda^{(k)}_i}{2\pi \iota})}{\prod_{l=1}^{k-1}\Gamma(1-\frac{\lambda^{(k-1)}_l-\lambda^{(k)}_i}{2\pi \iota})}\cdot \overline{m^{(k)}_i}.
\end{eqnarray*}
where $k=1,...,n-1$ and \[m^{(k)}_j=\sum_{j=1}^k \frac{(-1)^{k-j}\Delta^{1,...,\hat{j},...,k}_{1,...,k-1}(\lambda^{(k)}_i-A_{\infty})}{\prod_{l=1,l\ne i}^k(\lambda^{(k)}_i-\lambda^{(k)}_l)}a_{j,k+1}.\]
Here $\Delta^{1,...,\hat{j},...,k}_{1,...,k-1}(\lambda^{(k)}_i-A_\infty)$ is the $k-1$ by $k-1$ minor of the matrix $\lambda^{(k)}_i-A_{\infty}$ (the $\hat{j}$ means that the row index $j$ is omitted). 
Furthermore, the other entries are given by explicit algebraic combinations of the sub-diagonal elements.
\end{thm}

\begin{defi}\label{solcat-}
We denote by $\Phi_{-}(u;A_{-\infty})$ the solution of the isomonodromy equation \eqref{isoeq} with the asymptotics $A_{-\infty}$ at $u_{\rm cat}^{(-)}$ in the sense of \eqref{secondasy}.
\end{defi}
\begin{pro}\label{trans}
The Stokes matrices $S_\pm\big(u,\Phi_-(u;A_{-\infty})\big)$ satisfy
\begin{eqnarray}\label{+to-}
S_\pm\big(u,\Phi_-(u;A_{-\infty})\big)=P\cdot S_\mp\big(u,\Phi(u;PA_{-\infty} P)\big) \cdot P.
\end{eqnarray}
Here $P$ is the anti-diagonal $n\times n$ matrix that has $1$ as every anti-diagonal entry. 
\end{pro}
\pf If $F_\pm(z)\in {\rm GL}_n(\mathbb{C})$ are the canonical solutions of equation \eqref{Stokeseq} in ${\rm Sect}_\pm$, then one checks that the functions \begin{eqnarray}\label{Ftau}
F^\tau_-(z):= P\cdot F_-(-z)\cdot P \hspace{2mm} \text{and} \hspace{2mm} F^\tau_+(z):= P\cdot F_+(-z)\cdot P
\end{eqnarray}
are solutions of 
\begin{eqnarray}\label{pStokeseq}
\frac{dF}{dz}=\Big(-\iota PuP-\frac{1}{2\pi\iota}\frac{PAP}{z}\Big)\cdot F.
\end{eqnarray}
Following the asymptotics of $F_\pm$ at $z=\infty$ and the uniqueness of solutions of \eqref{pStokeseq} with prescribed asymptotics by Theorem \ref{uniformresum}, $F^\tau_-(z)$ and $F^\tau_+(z)$ are actually the canonical solutions of \eqref{pStokeseq} in the sectors ${\rm Sect}_+$ and ${\rm Sect}_-$ respectively.

By Definition \ref{defiStokes}, the Stokes matrices $S_\pm\Big(-PuP,PAP\Big)$ of \eqref{pStokeseq} are the transition matrices between $F^\tau_\pm(z)$.
Replacing $F^\tau_\pm(z)$ by $F_\pm(z)$ via the identity \eqref{Ftau}, we can then express the Stokes matrices of \eqref{pStokeseq} in terms of the transition matrices (the Stokes matrices $S_\pm(u,A)$ of \eqref{Stokeseq}) between $F_\pm(z)$. It follows that
\[S_\pm(u,A)=PS_\mp\Big(-PuP,PAP\Big)P, \]
once the correct defining sectors of $F^\tau_\pm(z)$ and the change in branch of ${\rm log}(z)$ in $z^{\frac{[A]}{2\pi \iota}}$ are accounted for.
The above identity holds for any $A\in\Herm(n)$, therefore for any fixed $u$ we can replace $A$ by $\Phi_-(u;A_{-\infty})$ to obtain
\begin{eqnarray}\label{id1}
S_\pm(u,\Phi_-(u;A_{-\infty}))=PS_\mp\Big(-PuP,P\Phi_-(u;A_{-\infty})P\Big)P.
\end{eqnarray}
Under the change of coordinates $v={\rm diag}(v_1,...,v_n)={\rm diag}(-u_n,...,-u_1)=-PuP$, the identity \eqref{id1} becomes (only the right hand side of \eqref{id1} is rewritten in the new coordinates $v$)
\begin{eqnarray}\label{id2}
S_\pm(u,\Phi_-(u;A_{-\infty}))=PS_\mp(v,P\Phi_-(v;A_{-\infty})P)P.
\end{eqnarray}
Note that $\Phi_-(u;A_{-\infty})$ is a solution of the isomonodromy equation, the left hand side, therefore the right hand side, of \eqref{id2} are constant (independent of $u$ and $v$ respectively). Then $P\Phi_-(v;A_{-\infty})P$ is also a solution of the isomonodromy equation \eqref{isoeq} in the new variables $v$ (that can also be verified directly).

The change of coordinates $v=-PuP$ brings the zone $u_{\rm cat}^{(-)}$ of $u$ to the zone $u_{\rm cat}^{(+)}$ of $v$. Then the given asymptotics $A_{-\infty}$ of $\Phi_{-}(u;A_{-\infty})$ at $u_{\rm cat}^{(-)}$ implies that the solution $P\Phi_-(v;A_{-\infty})P$ of \eqref{isoeq} has the asymptotics $PA_{-\infty}P$ as $v$ in the zone $u_{\rm cat}^{(+)}$. That is
\[P\Phi_-(v;A_{-\infty})P=\Phi(v;PA_{-\infty}P).\]
Together with \eqref{id2}, we get
\begin{eqnarray}\label{id3}
 S_\pm(u,\Phi_-(u;A_{-\infty}))=PS_\mp(v,\Phi(v;PA_{-\infty}P))P.
\end{eqnarray}
Since the right hand side of \eqref{id3} is independent of $v\in U_{\rm id}$, we see that the identity \eqref{id3} is equivalent to \eqref{+to-}. It finishes the proof.
\qed

\vspace{2mm}
Proposition \ref{trans}, together with Theorem \ref{mainthm}, gives the explicit expression of the Stokes matrices in terms of the asymptotics $A_{-\infty}$ at $u_{\rm cat}^{(-)}$. 
\subsection{Connection formula between the asymptotic zones $u_{\rm cat}^{(+)}$ and $u_{\rm cat}^{(-)}$}\label{conntwozones}
Recall that the asymptotic zones $u_{\rm cat}^{(+)}$ and $u_{\rm cat}^{(-)}$ are determined by \eqref{zone+} and \eqref{zone-}. Given a solution $\Phi(u)$ of the isomonodromy equation \eqref{isoeq}, we have computed the Stokes matrices as functions of the asymptotics parameter $A_\infty$ at $u_{\rm cat}^{(+)}$, and of the asymptotics $A_{-\infty}$ at $u_{\rm cat}^{(-)}$. As an immediate consequence, the knowledge of the Stokes matrices enables us to connect the two sets of asymptotics parameters.

\begin{thm}\label{1thm}
Assume that a solution $\Phi(u)\in\Herm(n)$ of the isomonodromy equation \eqref{isoeq} on $U_{\rm id}$ has 
\begin{itemize}
    \item the asymptotics $A_\infty$ at $u_{\rm cat}^{(+)}$;
    \item the asymptotics $A_{-\infty}$ at $u_{\rm cat}^{(-)}$,
\end{itemize}
then we have
\begin{eqnarray}\label{connform}
S_+(u, \Phi(u;A_\infty))=P S_-(u,\Phi(u;PA_{-\infty} P^{-1})) P^{-1}.
\end{eqnarray}
\end{thm}
\pf By the assumption of $\Phi(u)$ and the Definition \ref{solcat+} and \ref{solcat-}, we have $\Phi(u)=\Phi(u;A_\infty)=\Phi_-(u;A_{-\infty}).$
It gives
\[S_+(u, \Phi(u;A_\infty))=S_+(u,\Phi(u;A_{-\infty})).\]
Then replacing $S_+(u,\Phi(u;A_{-\infty}))$ by the right hand side of the identity \eqref{+to-} verifies the identity \eqref{connform}.
\qed

\vspace{2mm}
Theorem \ref{1thm} solves the connection problem between the two asymptotic zones of $\Phi(u)$. Indeed, the Stokes matrices give a complete set of independent first integrals of the isomonodromy equation, and they are explicitly expressed by the asymptotics of the same solution 
$\Phi(u)$ in two different zones. Thus the asymptotics $A_{-\infty}$ at $u_{\rm cat}^{(-)}$ is an implicit function of the asymptotics $A_{\infty}$ at $u_{\rm cat}^{(+)}$ by the implicit equation \eqref{connform} (independent of $u$). By Theorem \ref{isomonopro}, the implicit equation \eqref{connform} can be explicitly expressed by $A_{\infty}$ and $A_{-\infty}$ involving gamma functions, eigenvalue functions and so on. Therefore, the connection formula of the higher rank analog of Painlev\'{e} VI, though complicated, is still explicit.

\subsection{The connection problem for generic asymptotic zones}
As we have seen from the previous subsections, in the study of the connection problem via the method of isomonodromic deformation, the key point is to find the expression of the Stokes matrices $S_\pm(u,\Phi(u))$ by the asymptotics of the solution $\Phi(u)$ of the isomonodormy equation in an asymptotic zone $u_{\rm asy}$. Actually, following \cite{Xu}, a generalization of Theorem \ref{mainthm} from $u_{\rm cat}^{(+)}$ to any other asymptotic zone $u_{\rm asy}$ is possible: it consists in reducing the computation of Stokes matrices of \eqref{isoStokeseq} to the computation of the monodromy data of the equations taking the form (the so-called analytic branching rule in \cite{Xu})
\begin{eqnarray}\label{branch}
\frac{dF}{dz}=\Big(\iota D_k-\frac{1}{2\pi\iota}\frac{A}{z}\Big)\cdot F,
\end{eqnarray}
where $F(z)$ is valued in ${\rm GL}_n(\mathbb{C})$, $A\in{\Herm}(n)$ and \[D_k={{\rm diag}(\underbrace{0,...,0}_{k}, \underbrace{1,...,1}_{n-k})}, \ \text{for} \ k=1,...,n-1.\]

Roughly speaking, if the irregular term $u$ of the system \eqref{isoStokeseq} depends on a real parameter $t$ in the way that $u(t)={\rm diag}(u_1,...,u_{k}, t+u_{k+1},...,t+u_{n})$, then the equation \eqref{branch} was used in \cite{Xu} to decompose (or "branch") the $n\times n$ system \eqref{isoStokeseq} to the upper left and lower right subsystems of rank $k$ and $n-k$ respectively, as $t\rightarrow+\infty$. One sees that along the way the asymptotics of $\Phi(u(t))$ as $t\rightarrow\infty$ naturally arises.
One can imagine that the $t\rightarrow+\infty$ limit in $u(t)$ corresponds to a branching of a rooted tree with $k$ leaves on the left branch and $n-k$ leaves on the right branch. Then to get the expression of Stokes matrices of \eqref{isoStokeseq} via the asymptotics of $\Phi(u)$ at $u_{\rm asy}$, we just iterate the step according to the branching at each of the inner vertex of the rooted planar binary tree representing $u_{\rm asy}$. For example, if the asymptotic zone $u_{\rm asy}$ is given by Figure 1, we only need to compute the monodromy data for the type of equation \eqref{branch} with six different pairs $(n,k)$:
\[\left\{
          \begin{array}{lr}
             (n=6, k=2) \ \text{corresponding to the root vertex}  \\
           (n=2, k=1) \ \text{corresponding to the inner vertex with leaves $u_1$ and $u_2$}  \\
           (n=4, k=1) \ \text{corresponding to the inner vertex with leaves $u_3$, $u_4$, $u_5,u_6$}\\
           (n=3, k=2) \ \text{corresponding to the inner vertex with leaves $u_4$, $u_5,u_6$}\\
           (n=2, k=1) \ \text{corresponding to the inner vertex with leaves $u_4$, $u_5$}
             \end{array}
\right.\]

As $k=n-1$ or $k=1$, the equation \eqref{branch} is solved by the confluent hypergeometric functions. Thanks to the known global analysis of the functions, the monodromy of \eqref{branch} then has a closed formula. According to the analytic branching rule, a manipulation of the iteration of the computation for the equation \eqref{branch} with $(n=m,k=m-1)$ and $m=2,...,n$ eventually gives Theorem \ref{mainthm}. See \cite{Xu} for more details.

Although the equation \eqref{branch} for $2\le k\le n-2$ becomes harder to solve, its monodromy can be understood via the (classical analog of) representations of the Yangian of $\gl_k$. See our paper \cite{TX}. As discussed above, iterating the computation can express the Stokes matrices $S_\pm(u,\Phi(u))$ by the asymptotics of $\Phi(u)$ at any asymptotic zone. It thus helps to answer the connection problem $(b)$ of \eqref{isoeq} between all asymptotic zones in an explicit term. 

In the end, we would like to remark that the connection problem considered in this paper can be characterized by the Gelfand-Tsetlin theory in representation theory. And the explicit "algebraic characterization" of the analytic properties has much to do with the (quantization of) Hamiltonian description of the isomonodromy equation \cite{BoalchG, Re}.

\begin{rmk}
The connection formula in Theorem \ref{1thm} connects the asymptotics of solutions of \eqref{isoeq} at two different asymptotic zones. There is another type of connection formula, called the wall-crossing type formula in \cite{Xu}, which connects the asymptotics of solutions of \eqref{isoeq} as $u$ crosses a common face of the closure of two different connected components of $\h_{\rm reg}(\mathbb{R})$ in the De Concini-Procesi space. The latter detects the ($S_n$-equivalent) fundamental group of the De Concini-Procesi space of $\h_{\rm reg}(\mathbb{R})$, and can be used to give a monodromy realization of the cactus group action on the Gelfand-Tsetlin cone introduced by Berenstein and Kirillov \cite{BK}.
\end{rmk}

\subsection*{Acknowledgements}
\noindent We would like to thank Davide Guzzetti for useful comments. The author is supported by the National Key Research and Development Program of China (No. 2021YFA1002000) and by the National Natural Science Foundation of China (No. 12171006).

\Addresses
\end{document}